\pdfoutput=1
\documentclass[10pt,conference,compsocconf]{IEEEtran}
\usepackage[pdftex]{graphicx} 
\usepackage{color}
\usepackage{pdfsync} % comment out if not using TeXShop
\usepackage{hyperref}
\usepackage{cite}
\ifCLASSINFOpdf
\else
\fi
\usepackage{array}
\usepackage[caption=false,font=footnotesize]{subfig}
\usepackage{stfloats}
\usepackage{url}
\begin{document}
\title{\Large\bf Subband coding for large-scale scientific simulation data using {JPEG 2000}}


% This is crap; it indents the author block in the left column so much it crosses the center gutter on the page:

%\author{\IEEEauthorblockN{Christopher M. Brislawn, Jonathan L. Woodring,\\
%Susan M. Mniszewski, and James P. Ahrens}
%\IEEEauthorblockA{Computer, Computational \& Statistical Sciences Div.\\
%Los Alamos National Laboratory\\
%Los Alamos, NM 87545--1663 USA\\
%Email: \{brislawn,woodring,smm,ahrens\}@lanl.gov}
%\and
%\IEEEauthorblockN{David E. DeMarle}
%\IEEEauthorblockA{Kitware, Inc.\\
%21 Corporate Drive\\
%Clifton Park, NY 12065--8662 USA\\
%Email: dave.demarle@kitware.com}
%}


\author{\IEEEauthorblockN{Christopher M. Brislawn\IEEEauthorrefmark{1},
Jonathan L. Woodring\IEEEauthorrefmark{1 },
Susan M. Mniszewski\IEEEauthorrefmark{1},\\
David E. DeMarle\IEEEauthorrefmark{2},
and James P. Ahrens\IEEEauthorrefmark{1}
}
\IEEEauthorblockA{\IEEEauthorrefmark{1}Los Alamos National Laboratory, Los Alamos, NM 87545--1663 USA\\
Email: \{brislawn,woodring,smm,ahrens\}@lanl.gov}
\IEEEauthorblockA{\IEEEauthorrefmark{2}Kitware, Inc., 21 Corporate Drive, Clifton Park, NY 12065--8662 USA\\
Email: dave.demarle@kitware.com}
}

\IEEEspecialpapernotice{(Invited Paper)}

\maketitle

\begin{abstract}
The ISO/IEC JPEG~2000 image coding standard is a family of source coding algorithms targeting high-resolution image communications. JPEG 2000 features highly scalable embedded coding features that allow one to interactively zoom out to reduced resolution thumbnails of enormous data sets or to zoom in on highly localized regions of interest with very economical communications and rendering requirements. While intended for fixed-precision input data, the implementation of the irreversible version of the standard is often done internally in floating point arithmetic. Moreover, the standard is designed to support high-bit-depth data. Part 2 of the standard also provides support for three-dimensional data sets such as multicomponent or volumetric imagery. These features  make JPEG~2000  an appealing candidate for highly scalable communications coding and visualization of two- and three-dimensional data produced by scientific simulation software. We present results of initial experiments  applying JPEG~2000  to scientific simulation data produced by the Parallel Ocean Program (POP) global ocean circulation model, highlighting both the promise and the many challenges this approach holds for scientific visualization applications.
\end{abstract}


\section{Introduction}
\label{Intro}
\begin{figure}[t]
\centering
\includegraphics[width=3.375in]{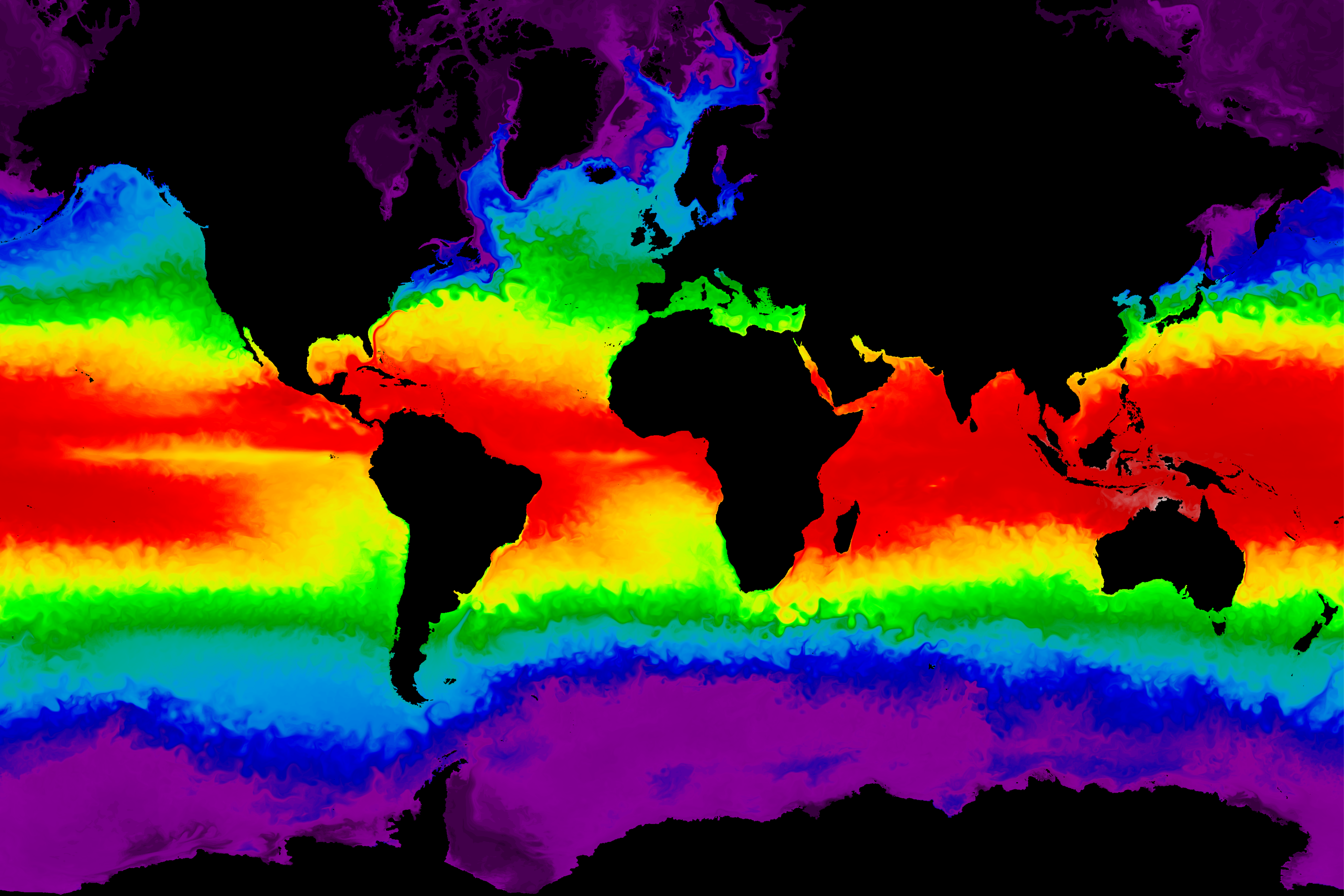}
\caption{Surface temperature field from a POP ocean  simulation.}
\label{fig:eco_temp41}
\vspace*{-0.375in}
\end{figure}

While  scientific computing  continues to tackle bigger and bigger  modeling and simulation problems (e.g., Figure~\ref{fig:eco_temp41}), improvements in serial hardware performance have slowed  as clock rates have stagnated at a few GHz. To maintain a steady increase in supercomputing performance, the recent trend  has been an exponential growth in processor parallelism. This is  changing the fundamental bottlenecks in high-performance computing (HPC). When supercomputers reach exascale ($10^{18}$ operations/second) some time in the next decade, the  limiting factor in system performance is expected to be power consumption for data movement and memory, not floating point computations. It will  be impossible to move all of the data computed by an exascale simulation out of core and into  nonvolatile storage fast enough to maintain a reasonable pace for the simulation.  For a good overview of the prospects for exascale computing, see the reports at
\centerline{\footnotesize    http://science.energy.gov/ascr/news-and-resources/program-documents}

In such a bandwidth-limited HPC environment, it is imperative to make the best possible use of the available communications bandwidth, memory, and nonvolatile storage. This means that source coding and compression tailored for simulation data will likely play an important role in successful exascale HPC data management and analysis. Source coding represents a new direction for scientific computing, however, since communications issues in large-scale scientific computing have traditionally been addressed with  hardware solutions, an approach that is not expected to scale up to exascale.

Adding a new core capability to the HPC technology mix raises many new questions and challenges. \emph{How} should data be encoded for optimal HPC bandwidth utilization? \emph{Where} in the simulation  flow should encoding be performed? \emph{Which part} of the data is most amenable to significant  bandwidth reduction? \emph{What effects} will bandwidth reduction have on data analysis and visualization? \emph{How much entropy} needs to be retained in the various components of an HPC data stream to preserve the scientific answers  the simulation was designed to deliver? 

This paper presents  recent efforts by the authors to start addressing some of these questions, primarily in the context of scientific visualization, using the JPEG~2000  image coding standards. JPEG~2000 is certainly not the only source coding approach applicable to HPC  simulation data. Some tools, notably VAPOR~\cite{RastClyne2008},  simply store uncompressed floating point data in a  multiresolution framework produced by a wavelet transform decomposition of the data. Older approaches~\cite{BradBris93b,BBQZN95,trott_96,ihm1998,kim1999,rodler1999,guthe2002,wang2005} compressed wavelet transform coefficients using combinations of coefficient thresholding or quantization, run-length coding, and/or Huffman coding. These older approaches are gradually being superseded, however, by modern embedded coding techniques that offer greater scalability, better bandwidth efficiency, fine-grained rate control, and decoder-driven progressive transmission capabilities. The ability to take advantage of these state-of-the-art source coding features using a standards-based toolkit  makes JPEG~2000 an attractive candidate for HPC data management.

\section{JPEG 2000 Coding}
\label{J2K}
\begin{figure}[tb]
\centering
\includegraphics[width=3.375in]{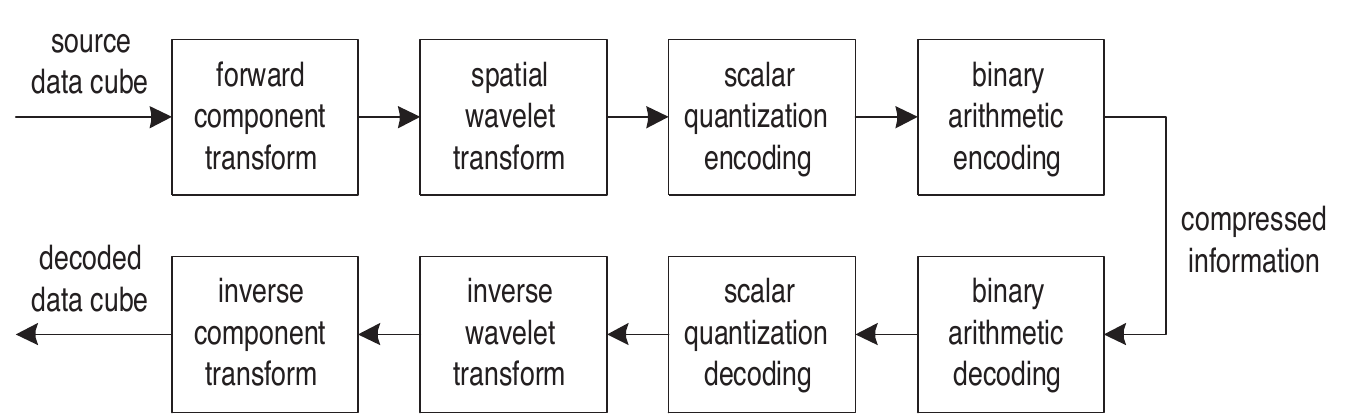}
\caption{Overview of JPEG 2000 irreversible encoding and decoding.}
\label{fig:J2K_overview}
\vspace*{-0.39in}
\end{figure}
JPEG~2000~\cite{Usevitch:01:tutorial-modern-lossy,TaubMarc02,BrisQuirk03a,AcharyaTsai:04:JPEG2000-Standard} is a family of international standards for digital image coding developed by the Joint Photographic Experts Group (JPEG) of the International Organization for Standardization (ISO) and the International Electrotechnical Commission (IEC) and published jointly with the International Telecommunications Union (ITU)~\cite{ISO_15444_1,ISO_15444_2}. As shown in Figure~\ref{fig:J2K_overview}, the ``irreversible'' version of JPEG~2000 reads in logically rectangular source data and applies decorrelating transforms, usually in floating point arithmetic. In the HPC  context, the ``components'' in the input could correspond either to 2-D slices in volumetric data or to multiple 2-D physical fields. The decorrelated subbands produced by cross-component and spatial wavelet transformation are then quantized to fixed-point values and entropy-encoded  using block-based binary arithmetic bit-plane coding. (An alternative ``reversible''  path transforms fixed-point input using nonlinear  integer-to-integer wavelet transforms, bypassing the quantization step and  enabling lossless subband coding.)

This encoding process makes the JPEG~2000 representation highly local and highly scalable with respect to a number of important parameters. Wavelet transform coefficients produced by short FIR filters are spatially localized samplings of the input data, and the hierarchical nature of wavelet transform decompositions makes JPEG~2000 intrinsically scalable with respect to spatial resolution. The block-based structure of the arithmetic encoding ensures that the compressed output preserves the spatial localization inherent in the wavelet transform coefficients.  The state-based binary arithmetic bit-plane encoding also makes the JPEG~2000 representation highly scalable with respect to sample precision: approximate values for the coefficients can be reconstructed in each coding block from an entropy-encoded bitstream truncated after any arbitrary number of coding passes.  

For the HPC data experiments conducted so far, the scalar quantization step has been designed using  subband quantization characteristics based on a nominal input entropy of around 25--27 bits per input sample (assuming 32-bit floating point data).  Entropy encoding then encodes \emph{all} quantized bit-planes, with ``quality layers'' set in the compressed codestream. Quality layering optimizes the reconstructed data decoded at a set of prespecified bit rates, typically spaced logarithmically; e.g., 8, 4, 2, 1, 0.5, 0.25 bits/sample.  This approach  results in relatively little (around 2:1) reduction in the total size in bytes of the input data but  restructures the data in a way that  enables \emph{huge} reductions in the bandwidth needed to reconstruct a region of interest at a  resolution and  precision appropriate for a particular  task, such as  visualization.  The ISO standard enables users to perform demand-driven, packet-level retrieval of compressed data  from a JPEG~2000 archive using the  JPEG~2000 Part~9 interactive client-server protocol (JPIP)~\cite{ISO_15444_9}. These  features---localization with respect to region of interest, spatial resolution, sample precision and data field, plus standardized client-server message queueing---totally decouple the cost of accessing small portions of a huge dataset from the size of the dataset. The client pays transmission and decoding/rendering costs only for the fraction of data that is explicitly requested.

\section{Analysis and Visualization of Compressed HPC Data}\label{Data}
\begin{figure}[t]
\centering
\includegraphics[width=3.375in]{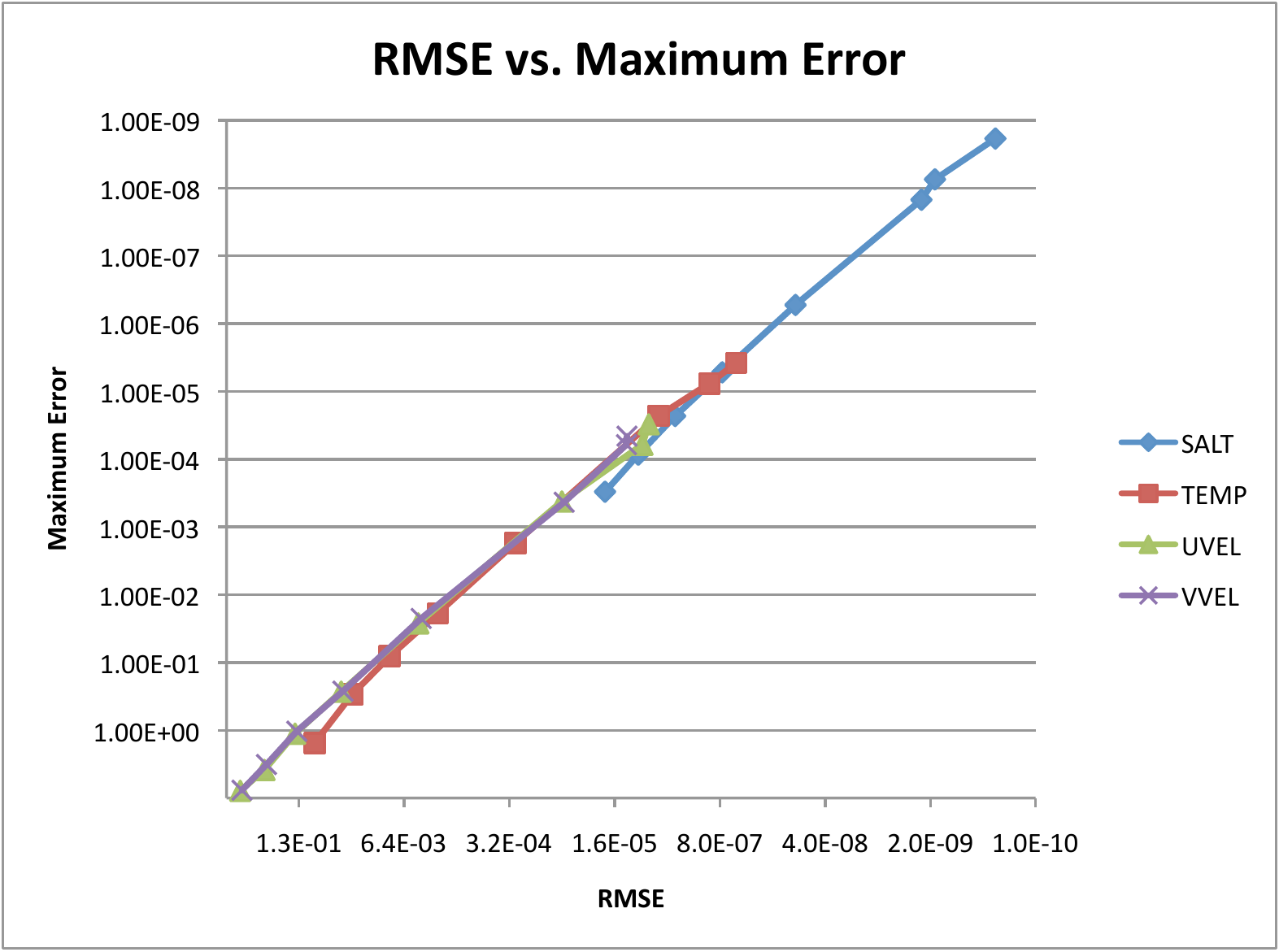}
\caption{Plots of max error vs. root-mean-square error for temperature, salinity, and velocity components at various bit rates.}
\label{fig:rmse_vs_maxerror}
\vspace*{-0.375in}
\end{figure}
The simulation shown in Figure~\ref{fig:eco_temp41} comes from  a global ocean circulation model with nominal $\frac{1}{10}^{\circ}$ resolution  generated by the Parallel Ocean Program (POP)~\cite{smith:lanltr02}. The $3600\times2400\times42$ arrays of  floating point data (temperature, salinity, E-W velocity, and N-S velocity) are  truncated to single precision for  scientific visualization and analysis, but this still amounts to 5.4~GB per time step. POP data is  compressed using the Kakadu JPEG~2000 software implementation per the above outline, saving all arithmetic coding passes in a compressed codestream with multiple quality layers. Kakadu~\cite{kakadu}, designed and written by  one of JPEG~2000's principal architects~\cite{Taubman:94:PhD-thesis}, is the most complete and most rigorously tested software implementation of JPEG~2000 available at present.  We note, though, that there is at least one open-source JPEG~2000 software project (OpenJPEG~\cite{OpenJPEGWebpage}) in active development. A JPEG~2000 reader has been implemented in ParaView~\cite{paraview} using the Kakadu library to enable visualization and various quantitative error analysis tasks for JPEG~2000-compressed data within ParaView/VTK. 

While Kakadu  maximizes the signal-to-noise ratio (SNR) in  images reconstructed at each quality layer bit rate, it is also desirable in HPC data management scenarios to quantify the maximum pointwise (or $L^{\infty}$) error in a reconstructed data set to provide scientific end-users with a worst-case pointwise error bound. Optimizing source coding schemes to minimize an $L^{\infty}$ rate-distortion metric is probably intractably hard, so it is of interest to see the highly linear relationship between RMSE and $L^{\infty}$ error reported in Figure~\ref{fig:rmse_vs_maxerror}. The $L^{\infty}$ error in Figure~\ref{fig:rmse_vs_maxerror} is proportional to RMSE for each physical field across reconstructed bit rates ranging from 8 down to 0.25 bits/sample. Particularly striking is the observation that the constant of proportionality is \emph{about the same} for all four fields,
\mbox{$L^{\infty}\mbox{\ error}\approx 10*\mbox{RMSE}$,} 
despite markedly different distributional properties for the different physical variables. This finding is very preliminary, but if it holds empirically in  more general contexts then it may be possible to model empirical $L^{\infty}$ error as a function of bit rate in terms of RMSE. 

\begin{figure}[b]
\vspace*{-0.25in}
\centering
\includegraphics[width=3.375in]{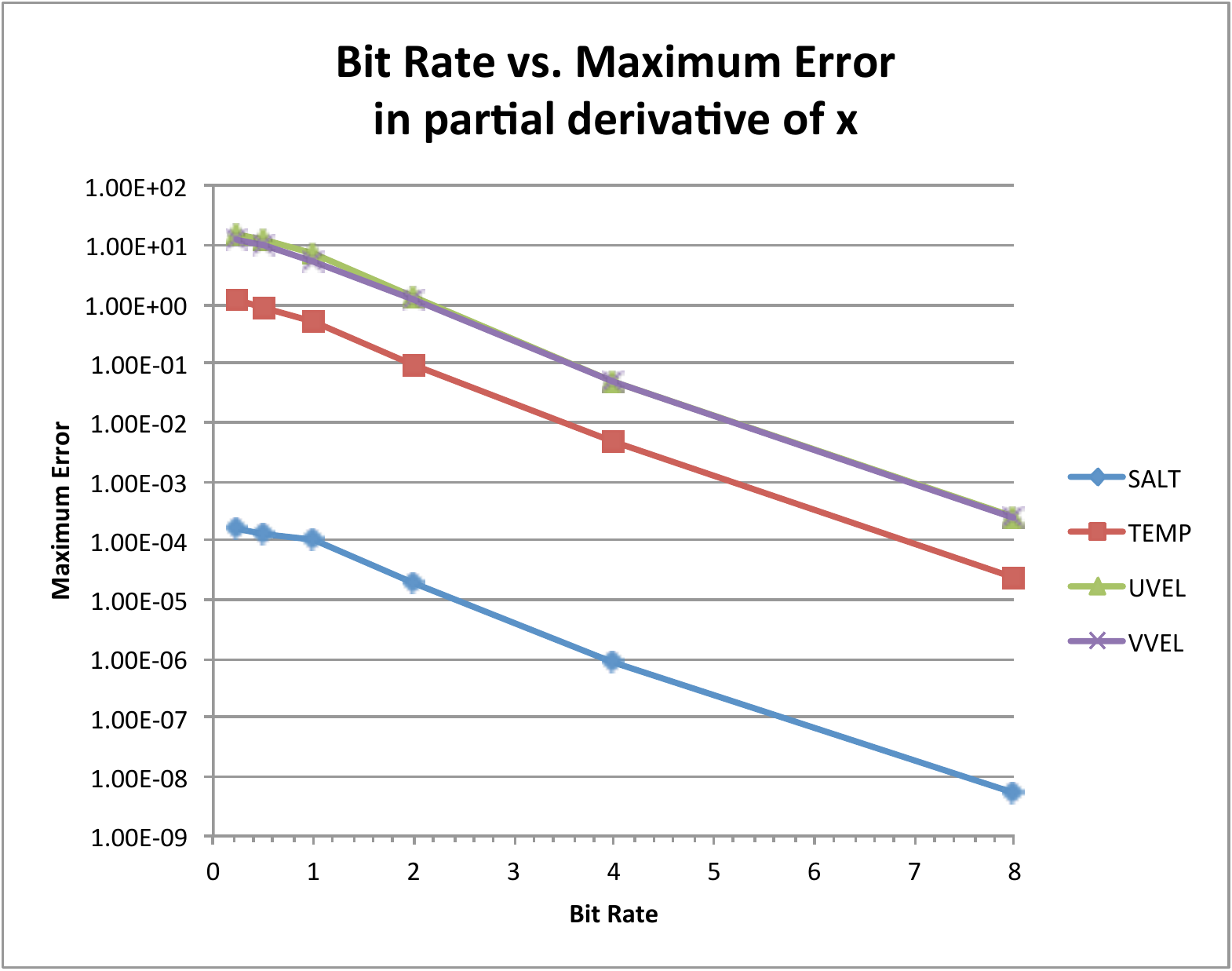}\label{fig:directional:EW}
\caption{Max error vs. rate for E-W directional derivatives of salinity, temperature, and velocity components at various bit rates.}
\label{fig:directional}
\end{figure}

In a similar preliminary vein, Figure~\ref{fig:directional} presents empirical $L^{\infty}$ rate-distortion behavior for numerical partial derivatives of the physical fields. Directional derivatives in the N-S direction behave similarly. Partial derivatives of velocity components are key features in the analysis of ocean eddies using the Okubo-Weiss approach~\cite{WHPSMAHH:11}. We expect SNR to have a certain range of linear rate-distortion behavior for well-designed source coders, but it is striking to see such linear $L^{\infty}$ rate-distortion behavior for numerical derivatives. Additional  rate-distortion analysis on JPEG~2000-encoded POP data is presented in~\cite{WoMnBrDeAh:11}.

\begin{figure}[t]
\centering
\subfloat[8 bits/pixel; max error = 1.49e-09]{\includegraphics[height=2.0in]{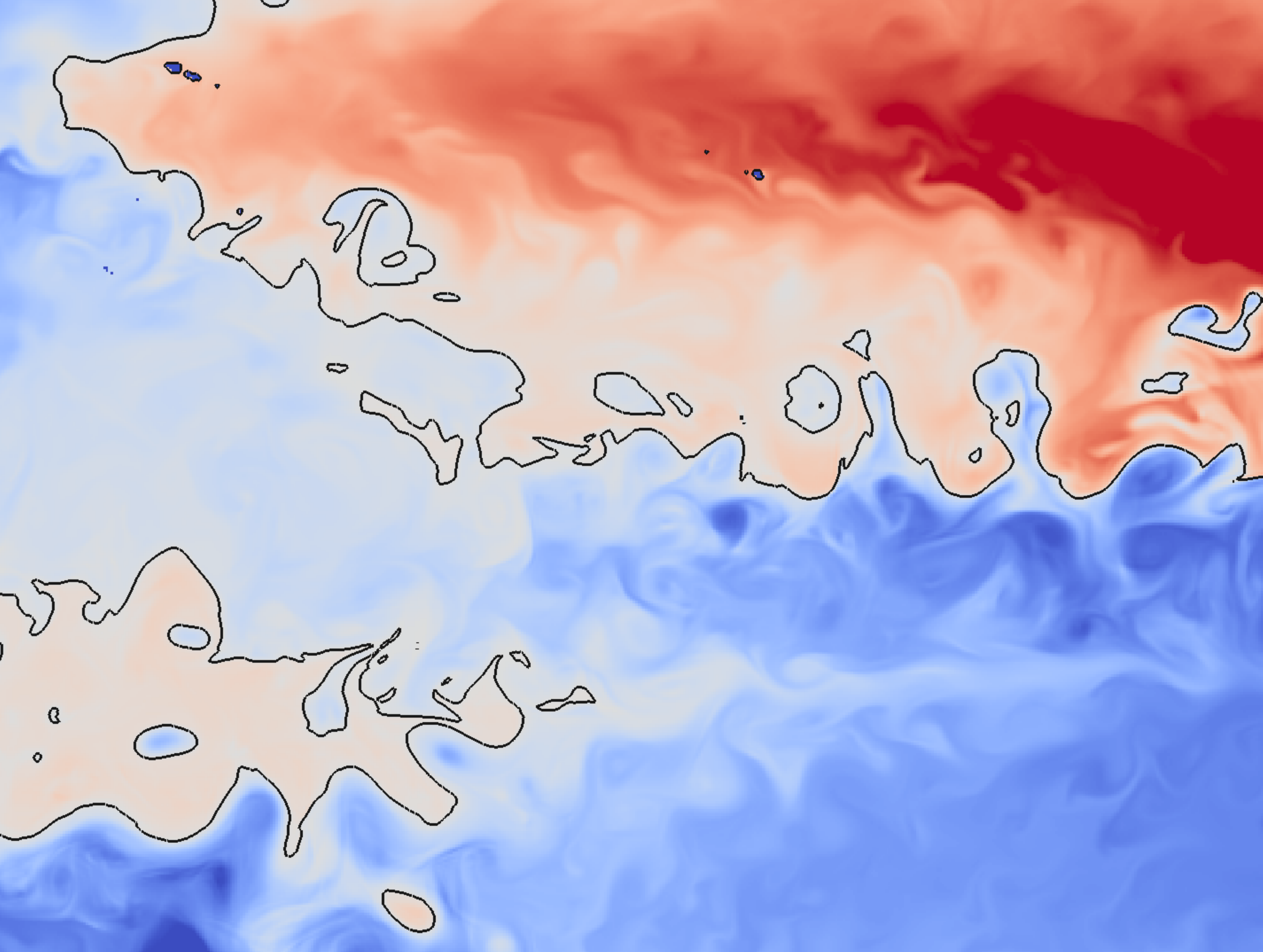}\label{fig:salt:8bpp}}\\
\vspace*{-1ex}
\subfloat[1 bit/pixel; max error = 2.31e-05]{\includegraphics[height=2.0in]{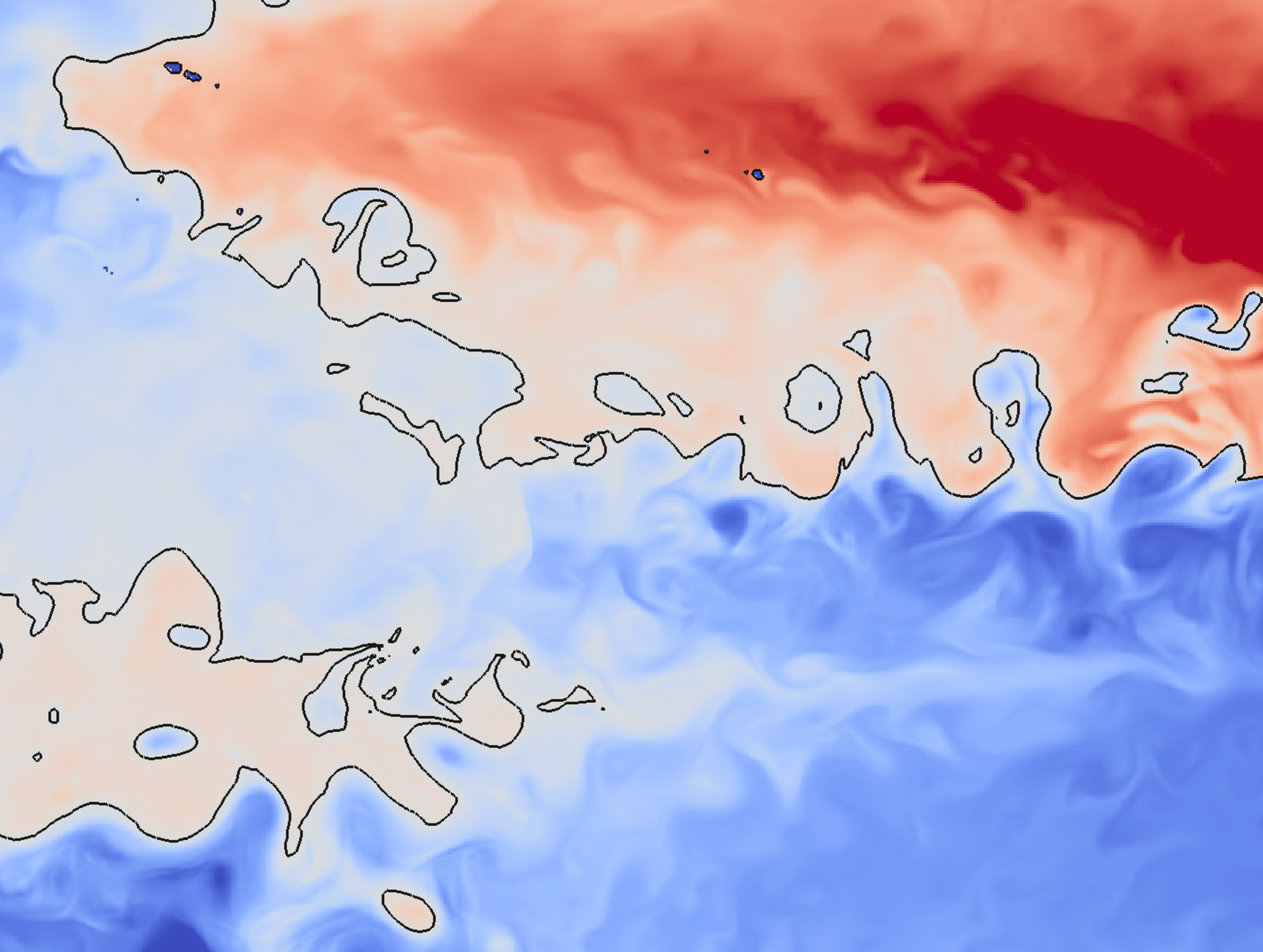}\label{fig:salt:1bpp}}\\
\vspace*{-1ex}
\subfloat[0.5 bits/pixel; max error = 8.59e-05]{\includegraphics[height=2.0in]{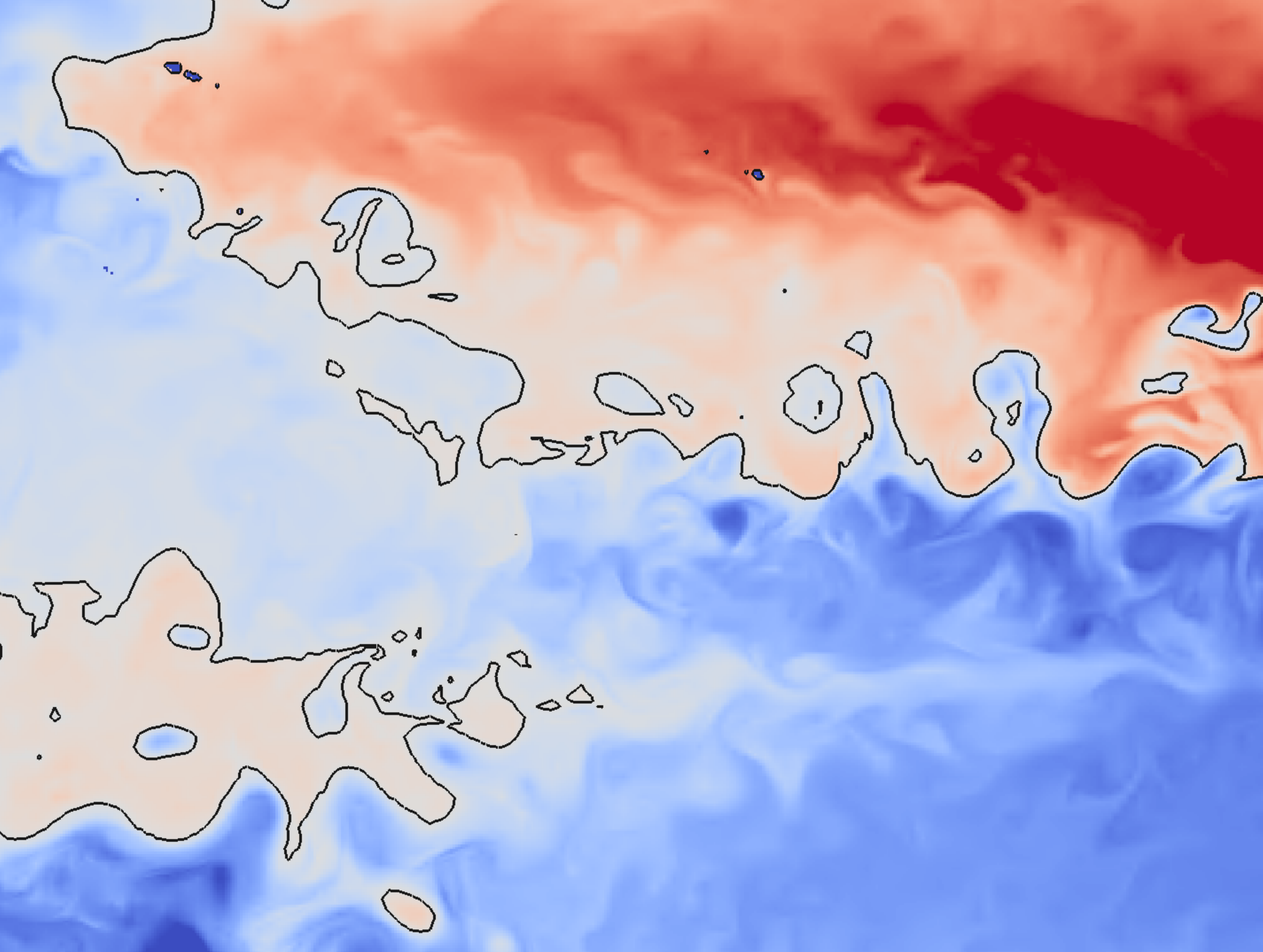}\label{fig:salt:p5bpp}}\\
\vspace*{-1ex}
\subfloat[0.25 bits/pixel; max error = 3.03e-04]{\includegraphics[height=2.0in]{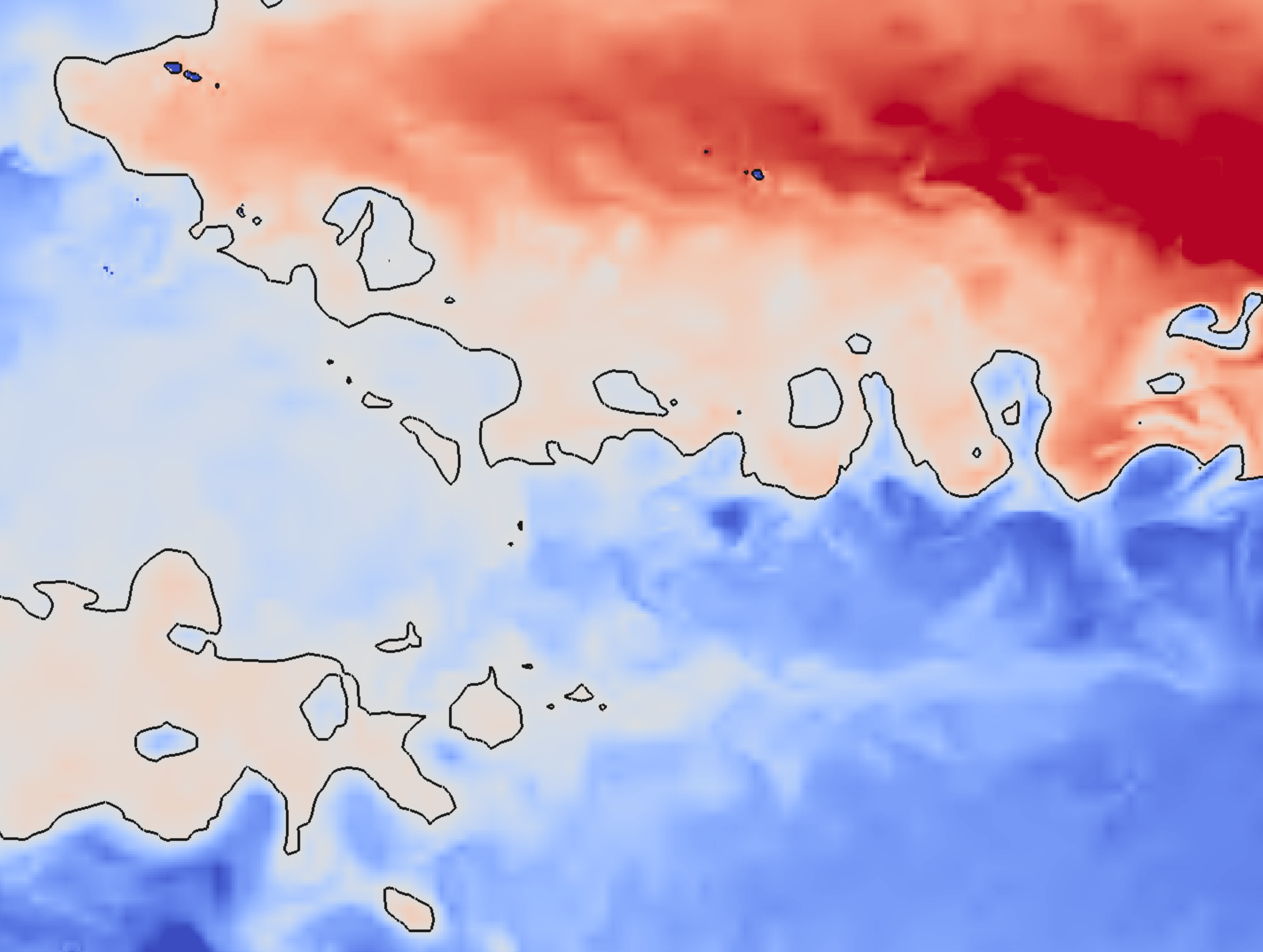}\label{fig:salt:p25bpp}}
%\vspace*{-1ex}
\caption{Isocontours on salinity data reconstructed at various bit rates.}
\label{fig:salt}
%\vspace*{-0.25in}
\end{figure}

To see the perceptual quality of  scientific visualization on compressed and reconstructed data,  Figure~\ref{fig:salt} shows a region of size $475\times358$ in a  salinity field progressively decoded and rendered using the ParaView JPEG~2000 reader at entropies of 0.25, 0.5,  1.0, and 8.0 bits/sample~\cite{WoMnBrDeAh:11}. Isocontours and amplitude colormaps were generated separately for each reconstructed array. The global maximum relative error (relative to the original 32-bit  input) in the  data reconstructed at 8.0 bits/sample is less than $10^{-7}$, or better than 140~dB SNR, which is roughly machine precision. The 8-bit reconstruction thus provides analyses and  visualizations that are indistinguishable from those obtained using the original 32-bit floating point data, in spite of being compressed 4:1. Small discrepancies, smoothing artifacts, and isocontour degradation are visible  at 0.5 and 0.25 bits/sample, but the degradation is graceful as rates drop below 1 bit/sample. It is thus entirely plausible that many basic scientific visualization tasks like these can be performed reliably at entropies as low as 1 bit/sample, a 32:1 reduction in bandwidth (1.5 orders of magnitude) relative to single-precision floating point. The potential savings from region-of-interest and reduced-resolution  retrieval from exascale data sets are even greater.

\section{Conclusions and Future Challenges}
\label{Concl}
The surface has barely been scratched when it comes to incorporating modern transform-based communications source coding standards in HPC data management. Part of the problem is that  HPC end-users traditionally have  not had to think much about bandwidth-efficient data management, and HPC suppliers have until now relied on big iron to overcome HPC bottlenecks. Consequently, there are no established best-practices in this area, so many HPC research groups have been experimenting with in-house floating point data compression schemes. The LANL-Kitware group is pursuing a path of standards-based source coding technology both to avoid reinventing as many wheels as possible and in recognition of the fact that HPC data encoding is really a \emph{communications} issue, and communications require the interested parties to agree on standardized protocols. In particular, it seems clear that \emph{someone} ought to explore just how far the JPEG~2000 standard can be pushed in the HPC arena.

There are plenty of open questions raised by this particular use of JPEG~2000.  Following are a few of the problems the LANL-Kitware group is working on (or  worrying about).
\begin{enumerate}
\item Efficiently estimate the entropy of floating point input to enable appropriate quantization of wavelet transform data in irreversible JPEG~2000 encoding processes.

\item Enable three-dimensional, multicomponent data encoding using JPEG~2000 Part 10~\cite{ISO_15444_10}.

\item Implement interactive multiscale client-server semantics between ParaView  clients and JPIP  servers.

\item Develop low-entropy  schemes for interpolating uninitialized grid regions in HPC data (e.g., continents and islands in POP ocean simulations).

\item In-situ  JPEG~2000 encoding for exascale simulations.
\end{enumerate}

\section*{Acknowledgments}
This work was supported by the Advanced Scientific Computing Research Program (ASCR) operated by the U.~S. Department of Energy's Office of Science.

%


%\newpage
\bibliographystyle{IEEEtran}
%\bibliography{CMBstring,CMBpubs,acad-press,elsevier,ieee,govt,maltzahn,math-soc,miscel,prentice-hall,spie,standards,CMBcrossref}
\bibliography{CMBstring,CMBpubs,SSIAI_12,standards,CMBcrossref}
\end{document}